\documentclass[nofootinbib,prb,amsmath,amssymb,twocolumn,floatfix]{revtex4-2}

\usepackage{graphicx}% Include figure files
\usepackage{dcolumn}% Align table columns on decimal point
\usepackage{bm}% bold math
\usepackage{siunitx}
\usepackage{mathtools}
\usepackage[hidelinks]{hyperref}% add hypertext capabilities
\usepackage{todonotes}
\graphicspath{ {images/} }

\renewcommand{\P}{\mathcal{P}}
\newcommand{\A}{\mathcal{A}}

\newcommand{\G}{\mathcal{G}}

\newcommand{\K}{\mathcal{K}}

\renewcommand{\P}{\mathcal{P}}
\newcommand{\R}{\mathbb{R}}

\renewcommand{\S}{\mathcal{S}}
\newcommand{\T}{\mathbb{T}}
\newcommand{\V}{\mathcal{V}}

\newcommand{\Z}{\mathbb{Z}}

\newcommand{\p}{\partial}
\newcommand{\dx}{\: \mathrm{d}}

\newcommand{\ds}{\displaystyle}
\newcommand{\iu}{\mathrm{i}\mkern1mu}

\newcommand{\uin}{u^\textrm{in}}
\newcommand{\usc}{u^\textrm{sc}}
\newcommand{\vin}{v^\textrm{in}}

\newcommand{\svdots}{\raisebox{3pt}{\scalebox{.6}{$\vdots$}}}
\newcommand{\sddots}{\raisebox{3pt}{\scalebox{.6}{$\ddots$}}}

\begin{document}

\preprint{APS/123-QED}

\title{Coupled harmonics due to time-modulated point scatterers}

\author{E.~O. Hiltunen$^{1}$ and B. Davies$^{2}$}
\affiliation{$^1$Department of Mathematics, University of Oslo, Moltke Moes vei 35, 0851 Oslo, Norway \\
$^2$Department of Mathematics, Imperial College London, London SW7 2AZ, United Kingdom}

\date{\today}

\begin{abstract}
We consider the resonance and scattering properties of a composite medium containing scatterers whose properties are modulated in time. When excited with an incident wave of a single frequency, the scattered field consists of a family of coupled harmonics at frequencies differing by the frequency of temporal modulation. Similarly, the temporal modulation induces coupling between the resonance frequencies, leading to exceptional points at certain modulation amplitudes. Moreover, the lack of energy conservation causes scattering coefficients to blow up when (complex) resonances cross the real axis. We have developed an integral operator approach to characterize the scattering problem and, for high-contrast scatterers, we present small-volume asymptotic formulas analogous to the classical results for the static (unmodulated) case. We conclude the paper with a boundary integral formulation of the time-modulated problem, which gives an efficient numerical approach and corroborates the asymptotic formulas.
\end{abstract}

\maketitle

\section{Introduction}
Wave scattering from time-modulated materials, sometimes referred to as \emph{Floquet metamaterials}, has gained recent interest as the new frontier in metamaterial physics \cite{galiffi2022photonics}. Modulating materials in time as well as space offers an additional degree of freedom that can be exploited to achieve exotic effects, such as non-reciprocal transmission \cite{simon1960action, oliner1961wave, ourir2019active}, Fresnel drag \cite{huidobro2019fresnel}, frequency conversion \cite{yin2022floquet}, parametric amplification \cite{cullen1958travelling, raiford1974degenerate}, exceptional points \cite{nikzamir2022achieve,rouhi2020exceptional}, and topological localisation \cite{rechtsman2013photonic,fleury2016floquet}. Recent progress in the fabrication of time-modulated materials means such systems are now a reality, yielding an exciting new paradigm for wave control \cite{tirole2023double, peng2016experimental}.

Much of the literature in space-time metamaterials focuses on the case of synthetically moving gratings where the material parameters are modulated as functions of $x-c_g t$ for some uniform grating velocity $c_g>0$ \cite{pendry2021gain}. This setting simplifies the mathematical analysis, since a transformation into a moving coordinate frame allows explicit solutions to the scattering problem \cite{lurie2007introduction}. In these situations, a homogenization theory has been developed \cite{huidobro2019fresnel, nassar2017modulated, huidobro2021homogenization, touboul2023high}. Although more general time dependencies have been studied in several situations, including effective gaps for Floquet Hamiltonians \cite{sagiv2022effective,hameedi2023radiative}, field patterns \cite{mattei2023effects}, time-dependent Mie scattering \cite{asadchy2022parametric}, and subwavelength resonance \cite{ammari2021time, feppon2022subwavelength, ammari2023transmission,ammari2024scattering}, the setting of general time modulation remains an open problem. In this work, we are interested in a composite medium of  temporally modulated scatterers, surrounded by a background medium that is both spatially and temporally constant. Specifically, we develop a mathematical framework for scatterers whose refractive index is periodically modulated in time. Due to the Floquet condition, the scattered field consists of multiple harmonics posed at frequencies differing by the modulation frequency. In close analogy to the static case, the system of coupled harmonics can be effectively solved using integral operator techniques.

\begin{figure}
	\centering
	\includegraphics{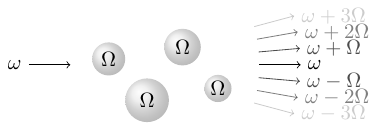}
	\caption{An incident field of a single frequency $\omega$ will give rise to a family of scattered harmonics. If the scatterers are modulated at a frequency $\Omega$, the family of coupled harmonics will occur at frequencies $\omega+n\Omega$ for integer $n$.}\label{fig:sketch}
\end{figure}

A common principle in the literature on time-modulated materials is the duality between the spatial and temporal dimensions in the wave equation. For example, temporal interfaces will give rise to ``time reflection'' and ``time refraction'' \cite{mendoncca2002time,koutserimpas2018electromagnetic}. As another example of this principle, time-modulated systems may exhibit a ``momentum band structure'' and $k$-gaps, analogous to the frequency band structure and band gaps exhibited in spatially periodic media. However, in the current work, we consider wave scattering from compactly supported scatterers inside a static background with open boundary conditions, as sketched in Fig.~\ref{fig:sketch}. In this case, the ``time-space'' duality is broken by the outgoing radiation condition in the spatial far field \cite{feppon2022subwavelength}. As result, there is no associated momentum band structure, and the system instead exhibits scattering resonances. These resonances arise as poles of the resolvent in the complex plane, analogously to resonances in the unmodulated case. Using integral operators, we can succinctly account for the outwards-propagating nature of the scattered field by using the outgoing Green's function, and the resonances satisfy a nonlinear eigenvalue problem for the integral operator.

In this work, we mainly consider the resonant case when the refractive index (in three dimensions) scales as $O(\epsilon^{-2})$, where $\epsilon$ is the diameter of each scatterer. Under this scaling, the cavity modes of each scatterer couple with the incident field, causing a strongly scattered field at certain resonant frequencies of the incident field. As a consequence of the coupled harmonics due to time modulation, these resonant frequencies occur periodically along the frequency axis, and will be called resonant \emph{quasi}frequencies. We derive formal asymptotic expansions of the resonant quasifrequencies of time-modulated high-contrast scatterers. This way, we are able to reduce the $3+1$-dimensional wave PDE model into an ODE in the time variable, thereby greatly reducing the computational complexity of computing the resonances.

Time-modulated systems typically exhibit frequency conversion, whereby the frequency of the scattered field is different to the frequency of the incident field. More specifically, a single incident frequency gives rise to a family of scattered harmonics whose frequencies differ by multiples of the frequency of time modulation. In the point-scattering approximation, we therefore have an (infinite) matrix of scattering coefficients, where the $(i,j)$\textsuperscript{th} entry describes the amplitude of the $i$\textsuperscript{th} scattered harmonic resulting from the $j$\textsuperscript{th} incident harmonic. For frequencies close to resonance, we use a modal decomposition to find explicit formulas for these scattering coefficients.

In the static case (where the refractive index is constant in $t$), energy conservation forces eigenmodes of real resonances to be bound states which do not couple to the spatial far field. By contrast, energy conservation is broken in time-modulated systems. In general, real resonances will now correspond to extended states which can be excited by an incident plane wave. Around such parameter points, the time modulation serves to inject energy into the scattered field, resulting in a blowup of the scattering coefficients and a strongly scattered field. Based on the asymptotics of the resonances, we demonstrate points where the resonances become real, and show that the scattering coefficients indeed blows up around these points.

\section{Problem formulation}\label{sec:vol}

We study the scalar wave equation
 in $d$ dimensions 
\begin{equation}\label{eq:wave}
	\left(\frac{\p^2}{\p t^2} - \frac{c^2}{n^2(x,t)}\Delta\right) u(x,t) = 0, \quad x\in \R^d, t\in \R.
\end{equation}
The refractive index $n(x,t)$ is supposed to depend on both $x$ and $t$, and is assumed to be $T$-periodic in $t$. We define the frequency $\Omega$ of time modulation as $\Omega = 2\pi/T$. Let $\uin$ be the incident field, which satisfies the free equation $(\p^2/\p t^2 -c^2\Delta)\uin(x,t)=0$. We assume that $\uin$ is a wave with quasifrequency $\omega$, in the sense that it can be written as
\begin{equation}
	\uin(x,t) = \sum_{n=-\infty}^\infty \vin_n(x) e^{-\iu (\omega+n\Omega)t}.
\end{equation}
Here, $\omega$ is defined in the (temporal) Brillouin zone $\T = \R/\Omega\Z$. Throughout, we will represent $\T$ by the first Brillouin zone $[-\frac{\Omega}{2}, \frac{\Omega}{2})$. The total field is a sum of the incident and scattered fields: $u = \uin + \usc$. We introduce the parameter $\eta$, which describes the modulation of $n^2$ relative to the constant background medium:
\begin{equation}
	1+\eta(x,t) = n^2(x,t).
\end{equation}
We seek solutions to \eqref{eq:wave} which are $\omega$-quasiperiodic in time $t$:
\begin{equation}
	u(x,t) = \uin(x,t) +  \sum_{n=-\infty}^\infty v_n(x) e^{-\iu (\omega+n\Omega)t}.
\end{equation}
This ansatz leads to a coupled system of equations for the coefficients $v_n(x)$, given by
\begin{equation}\label{eq:spatial}
\begin{split}
&\Delta v_n + k_n^2v_n + \sum_{m=-\infty}^\infty \eta_{n-m}(x) k_m^2 \left( v_m + \vin_m\right) =0.
%&\hspace{4cm} - \sum_{m=-\infty}^\infty \eta_{n-m}(x) k_m^2 \vin_m,
\end{split}
\end{equation}
where $k_n = (\omega+n\Omega)/c$ and $\eta_n(x)$ are the Fourier coefficients of $\eta(x,t)$, \emph{i.e.} 
\begin{equation}\eta(x,t) = \sum_{n=-\infty}^\infty \eta_n(x) e^{-\iu n\Omega t}.\end{equation}
We consider wave scattering from a system of finitely many heterogeneous scatterers embedded in a homogeneous background medium. In other words, we assume that $\eta$, as function of $x$, is constant and compactly supported on some domain $D\subset \R^d$. In this setting, the scattered field $v_n$ is outgoing in the sense that it can be represented in terms of the volume-integral operator $\G^k :L^2(D) \to L^2(D)$
\begin{equation}\G^k[\psi](x) = \int_D G^{k}(x-y)\psi(y) \dx y,\end{equation}
where $G^k$ is the outgoing Green's function for the Helmholtz equation, given by $G^k(x) = e^{\iu k|x|}/4\pi|x|$ in three dimensions. We then have the Lippmann--Schwinger form of equation \eqref{eq:wave}:
\begin{equation}\label{eq:main_sc}
	v_n - \sum_{m=-\infty}^\infty \eta_{n-m} k_m^2 \G^{k_n}[v_m+\vin_m] = 0,%-\sum_{m=-\infty}^\infty \eta_{n-m} k_m^2 \G^{k_n}[\vin_m].
\end{equation}
for all $n\in \Z$. Equation \eqref{eq:main_sc} takes the form of a system of coupled integral equations for each harmonic $v_n$. Resonances, or resonant quasifrequencies,  are $\omega \in \T + \iu\R$ for which there is a nontrivial solution to \eqref{eq:main_sc} with $\vin_m = 0$.

\section{Resonances of small high-contrast scatterers}\label{sec:asymp}
We consider how the resonances of a collection of scatterers behaves when the scatterers are small but highly contrasting (relative to the background medium). For simplicity, we will restrict to scattering in dimension $d=3$ for the remainder of this work, and outline the one-dimensional case in Appendix~\ref{sec:1d}. We suppose that the scatterers have characteristic size $\epsilon$, in the sense that we let $D=\epsilon B$ for some fixed domain $B$ and $0<\epsilon \ll 1$. In order to have resonance at finite frequencies as $\epsilon \to 0$, we will assume the scatterers are highly contrasting such that $\eta$ is proportional to $\epsilon^{-2}$. This setting can be viewed as a time-modulated generalization of the linear resonances characterized in \cite{meklachi2018asymptotic}. If we define the rescaling operator $P_\epsilon: L^2(D) \to L^2(B)$ as
\begin{equation}
	P_\epsilon[\phi](x) = \phi(\epsilon x),
\end{equation}
then we have that, as $\epsilon \to 0$, for fixed $k$,
\begin{equation} \label{prop:asymp}
	P_\epsilon \G^k (P_\epsilon)^{-1} = \epsilon^2\G^{(0)} + \epsilon^3\G^{(1)} + O(\epsilon^4),
\end{equation}
where the convergence is with respect to the operator norm in $L^2(B)$ and $\G^{(0)}, \G^{(1)}:L^2(B) \to L^2(B)$ are the operators
\begin{equation}
	\G^{(0)} = P_\epsilon\G^0(P_\epsilon)^{-1}, \quad \G^{(1)}[\phi](x) =  \frac{\iu k}{4\pi}\int_{B} \phi(y) \dx y.
\end{equation}
This follows directly from the asymptotic expansion
\begin{equation}
\epsilon G^k(\epsilon x) = \frac{1}{4\pi|x|} + \epsilon \frac{\iu k}{4\pi} + O(\epsilon^2),
\end{equation}
which is valid uniformly for $x$ in a compact domain $B$.

Using \eqref{prop:asymp}, we can now derive a formal asymptotic expansion for the resonant quasifrequencies as $\epsilon \to 0$. We begin with the case of a single scatterer. In the Lippmann--Schwinger integral equation \eqref{eq:main_sc}, we replace the volume-integral operator $\G^k$ by the asymptotic formula \eqref{prop:asymp} and neglect the terms that are pointwise $O(\epsilon^4)$ to yield the formula
\begin{equation}\label{eq:approx_formal}
	v_n(x) = \sum_{m=-\infty}^\infty \eta_{n-m}\epsilon^2 k_m^2\left( \G^{(0)} + \epsilon \G^{(1)}\right)[v_m](x).
\end{equation}
We can now diagonalize the problem in the spatial coordinate. Observe that $\G^{(0)}$ is a compact, positive, and self-adjoint operator. We let $(\mu_\ell, f_\ell)$ be the $L^2(B)$-normalized eigenvalue-eigenvector pairs of $\G^{(0)}$ and seek solutions to \eqref{eq:approx_formal} of the form
\begin{equation}\label{eq:phi}
	v_n(x) = a_n f_\ell(x) + \epsilon \hat v_n(x),\end{equation}
for some constants $a_n$ and some $\ell\in \Z$, where we assume $\hat v_n$ is some function such that $\langle f_\ell, \hat v_n\rangle = 0$ with respect to the $L^2(D)$-inner product. We can then pose \eqref{eq:approx_formal} in terms of the coefficients $a_n$, which reduces to 
\begin{equation}
	a_n = \left(\mu_\ell + \epsilon\frac{\iu k_nU_\ell^2}{4\pi} \right)\sum_{m=-\infty}^\infty \eta_{n-m}\epsilon^2 k_m^2a_m,
\end{equation}
where $U_\ell = \int_{B} f_\ell(x) \dx x$. Asymptotically as $\epsilon \to 0$, this is equivalent to 
\begin{equation}
	\left(1 - \epsilon\frac{\iu k_nU_\ell^2}{4\pi\mu_\ell} \right)a_n = \mu_\ell\sum_{m=-\infty}^\infty \eta_{n-m}\epsilon^2 k_m^2a_m.
\end{equation}
We can now transform the problem to the time domain. Define
\begin{equation}
	\psi(t) = \sum_{n=-\infty}^\infty a_ne^{-\iu (\omega+n\Omega)t}.
\end{equation}
In order to have resonances at finite frequencies as $\epsilon  \to 0$, we assume that
\begin{equation}
	\eta(t) = \frac{s(t)}{\epsilon^2}
\end{equation}
for some function $s(t)$ independent of $\epsilon$. This gives the boundary-value problem
\begin{equation}\label{eq:asymp_main}
	\begin{cases}
	\mu_\ell s(t) \psi''(t) + \epsilon c K_\ell\psi'(t)  + c^2\psi(t) = 0, \\[0.5em]
	\psi(t+T)=e^{-\iu \omega T} \psi(t),
	\end{cases}
\end{equation}
where $K_\ell = U_\ell^2/4\pi\mu_\ell$.

Equation \eqref{eq:asymp_main} is the main result of the asymptotic analysis. It is an eigenvalue problem for $\omega$, whose solutions approximate the resonant quasifrequencies of \eqref{eq:main_sc}.

In the case of a spherical scatterer, the first eigenvalue of $\G^{(0)}$ is $\mu_0 = 4\pi^{-2}$ while $K_0 = 8\pi^{-2}$. We therefore have the simplified asymptotic formula
\begin{equation}\label{eq:asymp_sphere}
	s(t) \psi''(t) + 2\epsilon c \psi'(t) + \frac{\pi^2c^2}{4}\psi(t) = 0.
\end{equation}

\begin{figure*}
	\centering
	\includegraphics{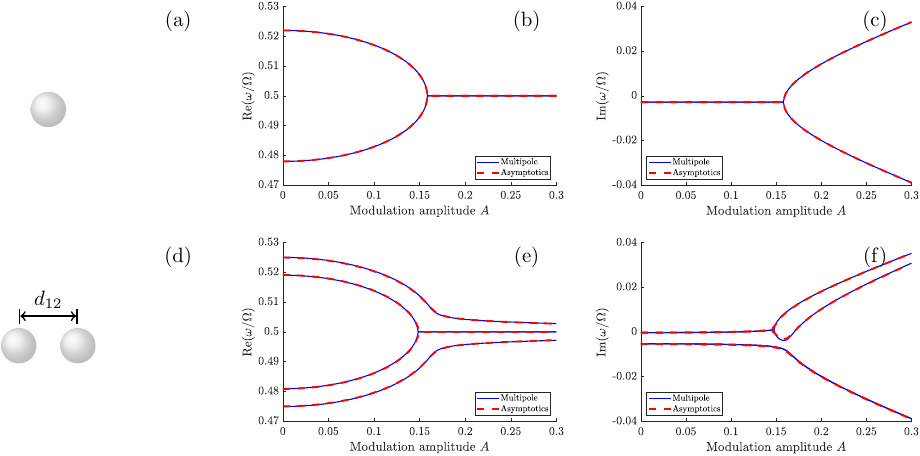}
	\caption{Computation of the resonant quasifrequencies as function of modulation strength $A$ in the case of spherical scatterers with cosine-modulation given by \eqref{eq:cosine}. These are computed either through a multipole discretisation of the full PDE (see Section~\ref{sec:BIE}) or by solving the asymptotic formula \eqref{eq:asymp_main}. We consider small scatterers with characteristic size $\epsilon = 0.01$ subjected to modulation with frequency $\Omega = 3$ and scaling parameter $s=1.2$. The real and imaginary parts of the resonant quasifrequencies for a single spherical scatterer are shown in (b) and (c). The scatter supports an exceptional point around $A\approx 0.15$. The resonant quasifrequencies for two spherical scatterers with separation distance $d_{12} = 0.5$ are shown in (e) and (f).} \label{fig:resonances}
\end{figure*}

For the examples in Fig.~\ref{fig:resonances}, we consider a cosine-modulation of the form
\begin{equation}\label{eq:cosine}
	\eta(t) = \frac{s}{\epsilon^2}\bigl(1 + A \cos(\Omega t)\bigr),
\end{equation}
where $A$ is the amplitude of modulation. We plot the resonant quasifrequencies computed through the asymptotic formula \eqref{eq:asymp_sphere} and compared against the boundary integral discretisation as outlined in Section~\ref{sec:BIE} below. Here we consider the case $\epsilon = 0.01$, whereby we observe close agreement between the asymptotics and the full system.

One important observation to make from \eqref{eq:asymp_main} is that the problem is not parity--time-symmetric due to the presence of the first-order derivative, originating from the radiation condition. Since this term has an $O(\epsilon)$ coefficient, it is parity--time-symmetric at leading order, which explains why the resonant quasifrequencies in Fig.~\ref{fig:resonances} are approximately either both real valued or in complex conjugate pairs. In other words, the leading order show the main characteristics of an exceptional point, as previously observed in time-modulated systems in \cite{nikzamir2022achieve,rouhi2020exceptional}. The presence of the first-order term, however, breaks this symmetry and is the reason that, for example, the imaginary part of both the resonant quasifrequencies is non-zero for small modulation amplitudes in Fig.~\ref{fig:resonances}(c).  

It is straightforward to repeat the analysis for multiple scatterers. We take $D$ to be a collection of $N$ scatterers centred at $z_i\in \R^3$ and separated by distance $d_{ij}$:
\begin{equation}D = \bigcup_{i=1}^N \epsilon B + z_i, \quad d_{ij} = |z_i-z_j|.\end{equation}
Letting $\eta(t) = \eta^i(t)$ in the $i$\textsuperscript{th} scatterer, and $s^i(t) = \epsilon^2\eta^i(t)$, we similarly derive the following system of boundary-value problems, for $i=1,...,N$,
\begin{equation}\label{eq:asymp_mainN}
	\begin{cases}
		\ds \mu_\ell s_i(t) \psi_i''(t)  + c^2\psi_i(t) \\ \ds \hspace{1.7cm}+ \epsilon c K_\ell\biggl(\psi_i'(t) - \sum\limits_{j\neq i}\frac{c}{d_{ij}}\psi_j\left(t-\tfrac{d_{ij}}{c}\right) \biggr) = 0, \\[0.5em]
		\psi_i(t+T)=e^{-\iu \omega T} \psi_i(t).
	\end{cases}
\end{equation}
In Fig.~\ref{fig:resonances}(d)--(f), we show the resonances of a system of two spheres computed through \eqref{eq:asymp_mainN}. Due to hybridisation of the scatterers, each resonance splits into two, and the asymptotic formula \eqref{eq:asymp_mainN} is able to closely describe this splitting.

\section{Point-particle scattering}\label{sec:scatter}
With the asymptotic analysis of the resonances at hand, we proceed by deriving the point-particle scattering approximation for incident frequencies $\omega$ close to these resonances. We take an incident field consisting of plane waves:
\begin{equation}\vin_j(x) = A_je^{\iu k_j\cdot x},\end{equation}
for incident amplitudes $A_j$, where $k_j$ is a wave vector of norm $\frac{|\omega+j\Omega|}{c}$. We let $D$ consist of a single scatterer centred at $z_0$: $D = \epsilon B + z_0$.
To proceed, we introduce the space  $\V = \ell^2\left(L^2(D)\right)$ of doubly infinite, square-summable sequences of functions in $L^2(D)$ under operator norm, and define an operator $\A(\omega)$ on $\V$ as follows:
\begin{equation}\A(\omega)\{v_n\}_{n\in \Z} = \left\{	\sum_{m=-\infty}^\infty \eta_{n-m} k_m^2 \G^{k_n}[v_m]\right\}_{n\in \Z}.\end{equation}
In this framework, a resonance $\omega$ of \eqref{eq:main_sc} corresponds to a pole of $\bigl(I - \A(\omega)\bigr)^{-1}$. Assume $\omega_0$ is a simple resonance corresponding to eigenmode $\Phi_0$ given as in \eqref{eq:phi}:
\begin{equation}\Phi_0 = \big\{a_nf_\ell(x)\big\}_{n\in \Z},\end{equation}
where we have neglected the small remainder $\epsilon \hat{v}_n$. For $\omega$ close to a simple resonance $\omega_0$, we have a Laurent expansion of the form
\begin{equation}\bigl(I-\A(\omega)\bigr)^{-1} = \frac{\P}{\omega-\omega_0} + \mathcal{R}(\omega),\end{equation}
where
\begin{equation}\P = \frac{\langle\Phi_0, \cdot \rangle }{\left\langle \Phi_0, \frac{\dx \A(\omega_0) }{\dx \omega} \Phi_0\right\rangle }\Phi_0,\end{equation}
and $\mathcal{R}(\omega)$ is holomorphic in a neighbourhood of $\omega_0$. Neglecting terms without the pole $(\omega-\omega_0)^{-1}$, we then have
\begin{equation}v_i(x) =  \frac{p}{\omega-\omega_0}\sum_{j=-\infty}^\infty \eta_{i-j} k_j^2 \G^{k_i}\left[\phi_j\right], \end{equation}
where
\begin{equation}p = \frac{\langle\Phi_0,\A(\omega)\{\vin_j\}_{j\in \Z} \rangle}{\left\langle \Phi_0, \frac{\dx \A(\omega_0) }{\dx \omega} \Phi_0\right\rangle }.\end{equation}
As $\epsilon \to 0$, we then have the far-field expansion of $v_i$, for $\omega$ close to $\omega_0$,
\begin{equation}\label{eq:sc_ff}
	v_i(x) = G^{k_i}(x-z_0)\frac{p}{\omega-\omega_0}\sum_{j=-\infty}^\infty \eta_{i-j} k_j^2 a_j \epsilon^3 U_\ell.
\end{equation}
Observe that, by linearity, we have
\begin{equation}p = \sum_{j=-\infty}^\infty p_j \vin_j(z_0),\end{equation}
where 
\begin{equation}p_j = \frac{\langle\Phi_0,\A(\omega)\{\delta_{jj'}e^{\iu k_j\cdot (x-z_0)}\}_{j'\in \Z} \rangle}{\left\langle \Phi_0, \frac{\dx \A(\omega_0) }{\dx \omega} \Phi_0\right\rangle }.\end{equation}
Asymptotically as $\epsilon \to 0$, we have
\begin{equation}
	p_j = \frac{c U_\ell k_j^2 \sum_{n\in \Z} \overline{a_n}\eta_{n-j}}{2\sum_{m,n\in \Z} \overline{a_n}a_m k_m \eta_{n-m}}.
\end{equation}
We can then define a matrix of scattering coefficients $\Lambda_{ij}$ for $i,j\in \Z$
\begin{equation}\label{eq:Lij}
	\Lambda_{ij} = 2\pi K_\ell\frac{\epsilon}{\omega-\omega_0}\frac{ a_ik_j^2\sum_{n\in \Z} \overline{a_n}\eta_{n-j}}{\sum_{m,n\in \Z} \overline{a_n}a_m k_m \eta_{n-m}},\end{equation}
which measures the amplitude of the scattered field of frequency $\omega+i\Omega$ given an incident field of frequency $\omega +j\Omega$.
In other words, for a general incident field $\uin(x,t) = \sum_{j=-\infty}^\infty \vin_j(x) e^{-\iu (\omega+j\Omega)t}$, the $i$\textsuperscript{th} mode of scattered field is given by
\begin{equation}\label{eq:scattered}
	v_i(x) = G^{k_i}(x-z_0) \sum_{j=-\infty}^\infty \Lambda_{ij}  \vin_j(z_0),\end{equation}
valid for small $\epsilon$ and for $\omega$ close to $\omega_0$. Equation~\eqref{eq:scattered} is the point-scattering approximation of the time-modulated scatterers. It shows how a single incident mode gives rise to a family of scattered modes at differing frequencies, where the coefficients $\Lambda_{ij}$ describe the frequency conversion.

\begin{figure*}
	\centering
	\includegraphics{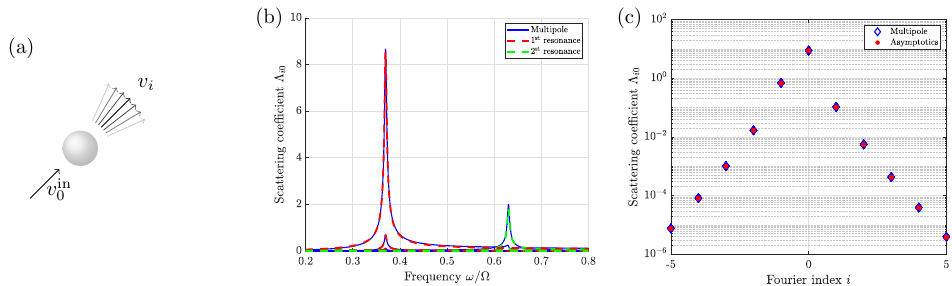}
	\caption{An incident field of a single frequency will give rise to a family of scattered harmonics, posed at frequencies which differ by multiples of the modulation frequency $\Omega$. Here, we compute the scattering coefficients of a single scatterer in the case of an incident field of frequency $\omega$, corresponding to $j=0$. We take a spherical scatterer of radius $\epsilon = 0.01$, with time modulation specified by $s=1.2$, $A=0.3$ and $\Omega = 4$. As shown in (b), the scatterer  supports two resonant peaks of the scattering coefficients. The asymptotics \eqref{eq:Lij} of the scattering coefficients are compared to the values computed through the multipole discretisation and are able to capture these resonant peaks. In (c), we show the scattering coefficients $\Lambda_{i0}$ at the first peak $\omega/\Omega \approx 0.38$, which decay exponentially in the Fourier index $i$.}\label{fig:scatter}
\end{figure*}

\begin{figure}
	\centering
	\includegraphics[width=0.8\linewidth]{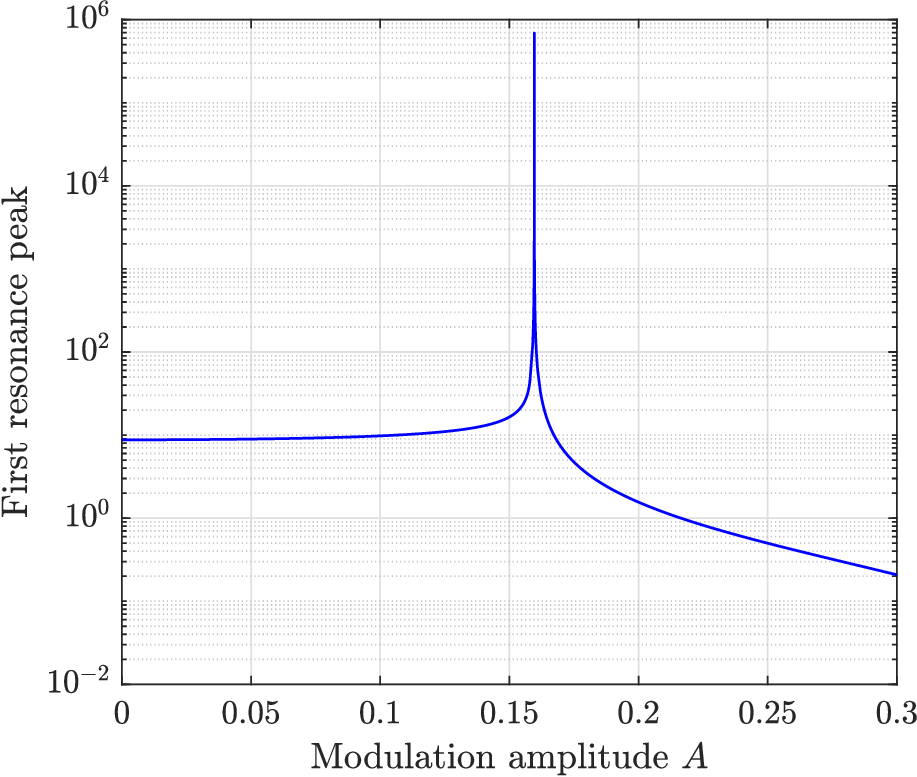}
	\caption{The scattering coefficients blow up when a (complex) resonance becomes real. Here, we plot the largest scattering coefficient as a function of modulation amplitude $A$. We use the same parameters as Fig.~\ref{fig:resonances}(b)--(c), whereby one of the resonances crosses the real axis around $A\approx 0.16$, resulting in a blowup of the scattering coefficients at this point.}\label{fig:blowup}
\end{figure}

\section{Numerical approach}\label{sec:BIE}
Since, at any given moment in time, $\eta$ is piecewise constant in space, the wave equation \eqref{eq:wave} can be solved straightforwardly using separation of variables in both the interior and exterior of the scatterers $D$. As we shall see, this will allow us to rephrase the volume integral equation of Section~\ref{sec:vol} into a boundary integral formulation (we refer, for example, to \cite{ammari2009layer} for a review of boundary integral techniques for wave resonance problems). Due to the reduction of dimensionality of the integrals, this formulation is better suited for numerical calculations. Truncating the resulting Fourier series yields a system of boundary integral equations which can be  straightforwardly solved with existing methods. This way, we numerically validate the formal asymptotics of Section~\ref{sec:asymp}--\ref{sec:scatter} in the case of spherical scatterers.  To ensure rapid convergence of the Fourier series, we require the time modulation $\eta(t)$ be smooth in $t$, whereby we observe exponential convergence of the method in terms of the length of the truncated Fourier series.

We consider a collection of $N$ scatterers $D^i$, and use the notation that inside $D^i$, $\eta(t)$ is equal to $\eta^i(t)$. Then,  let $\Phi_m^i(t)$ and $\lambda_m^i$ be the eigenfunctions and eigenvalues of the boundary value problem
\begin{equation} \label{eq:timeODE}
	\begin{cases}
		\ds \Phi''(t) + \frac{\lambda^2}{1+\eta^i(t)}\Phi(t) = 0,\\
		%	\Phi^i(t) \text{ is $\omega$-quasiperiodic}.
		\Phi(t+T)=e^{-\iu \omega T} \Phi(t).
	\end{cases}
\end{equation}
We can then write the solution to \eqref{eq:wave} as
\begin{equation} \label{eq:separable}
	u(x,t) = \begin{cases} 
		\ds\uin(x,t) + \sum_{n=-\infty}^\infty v_n(x)e^{-\iu(\omega+n\Omega)t}, & x \in \R^d\setminus D, \\
		\ds \sum_{m=-\infty}^\infty v_m^i(x) \Phi_m^i(t), & x\in D_i,
	\end{cases}
\end{equation}
where the spatial eigenmodes $v_n(x)$ in the background  $\R^d\setminus D$ and $v_m^i(x)$ in the interior of $D_i$ must satisfy the equations
\begin{equation} \label{eq:Helmholtzeqns}
	\begin{cases}
		\ds \Delta v_n + k_n^2 v_n = 0, \qquad &x \in \R^d\setminus D, \\
		\ds \Delta v_m^i + (k_m^i)^2 v_m^i = 0, \qquad &x \in D_i,
	\end{cases}
\end{equation}
where 
\begin{equation}k_n = \frac{\omega+n\Omega}{c}, \qquad k_m^i = \frac{\lambda_m^i}{c}.\end{equation}
The boundary conditions associated to \eqref{eq:Helmholtzeqns} should be chosen to guarantee the continuity of the field \eqref{eq:separable} and its normal derivative. 

A natural approach to solving the ordinary differential equation \eqref{eq:timeODE} is to express $\Phi^i$ as a Fourier series. Let $f_{mn}^i$ be the Fourier coefficients of $\Phi_m^i = \sum_{n=-\infty}^\infty f_{mn}^i e^{-\iu(\omega+n\Omega)t}$. Then, substitution into \eqref{eq:timeODE} yields the system of equations
\begin{equation} \label{eq:Fourier}
	f_{mn}^i(\omega+n\Omega)^2+\sum_{j=-\infty}^\infty \eta_{n-j}^i f_{mj}^i (\omega+j\Omega)^2 = (\lambda_m^i)^2 f_{mn}^i,
\end{equation}
where $\eta_n^i$ are the Fourier coefficients of $\eta^i(t) = \sum_{n=-\infty}^\infty \eta_n^i e^{-\iu n\Omega t}$. Once $f_{nm}$ have been obtained by solving \eqref{eq:Fourier}, the representation \eqref{eq:separable} for the solution $u(x,t)$ in the interior $x\in D_i$ can be written as 
\begin{equation}
	u(x,t) =  \sum_{n=-\infty}^\infty  \left( \sum_{m=-\infty}^\infty f_{mn}^i v_m^i(x) \right) e^{-\iu(\omega+n\Omega)t}.
\end{equation}
Then, it is clear that $u$ and its normal derivative will only be continuous across the boundary $\partial D$ at all times if
\begin{equation}\label{eq:bdycond}
	\begin{cases}
		\ds\sum_{m=-\infty}^\infty f_{mn}^i v_m^i \big|_- -  v_n\big|_+ = \vin_n,\\
		\ds \frac{\p}{\p\nu} \sum_{m=-\infty}^\infty f_{mn}^i v_m^i \big|_- -\frac{\p}{\p\nu} v_n\big|_+ = \frac{\p}{\p\nu} \vin_n,
	\end{cases}
\end{equation}
for all $n$. Here, $|_\pm$ denotes the limits from outside and inside $D$, respectively. This way, we have reduced the time-dependent wave equation to a system of Helmholtz equations \eqref{eq:Helmholtzeqns} in terms of $v_n$ and $v_n^i$, with associated boundary conditions \eqref{eq:bdycond}. We introduce the single-layer potential $\S_D^k$  as
\begin{equation}\S_D^k[\psi](x) = \int_{\partial D} G^{k}(x-y)\psi(y) \dx \sigma(y).\end{equation}
This way, we can represent the solution to \eqref{eq:Helmholtzeqns} as
\begin{equation}v_n = \S_D^{k_n}[\phi_n], \qquad v_n^i = \S_{D_i}^{k_n^i}[\psi_n^i],\end{equation}
where the unknown variables are now $\phi_n$ and $\psi_n^i$. The equation \eqref{eq:wave} is then equivalent to the following system of boundary integral equations: 
\begin{equation}\label{eq:bdy}
	\begin{cases}
		\ds\sum_{m=-\infty}^\infty f_{mn}^i \S_{D_i}^{k_m^i}[\psi_m^i] -\S_D^{k_n}[\phi_n] = \vin_n,\\
		\!\begin{aligned}\ds  \sum_{m=-\infty}^\infty f_{mn}^i & \left(\frac{1}{2}I  +\K_{D_i}^{k_m^i,*}\right)[\psi_m^i] \\ \ds &- \left(-\frac{1}{2}I  +\K_D^{k_n,*}\right)[\phi_n] = \frac{\p}{\p\nu} \vin_n,\end{aligned}\end{cases}\end{equation}
for all $i = 1,..., N$ and $n\in \Z.$ Here, $\K_{D}^{k,*}$ denotes the Neumann-Poicar{\'e} operator, defined by 
\begin{equation}
	\K_D^{k,*}[\phi](x) = \int_{\partial D} \frac{\partial }{\partial \nu_x}G^k(x-y) \phi(y) \dx \sigma(y),
\end{equation}	
for $ x \in \partial D$. In the case $D$ consists of a union of spheres, we can discretise $\S_D^k$ and $\K_D^{k,*}$ using a truncated basis of spherical harmonics as outlined in \cite[Appendix A]{ammari2020topologically}. We refer to this numerical approach as the ``multipole method'' and have used it to compute the numerical solutions in Fig.~\ref{fig:resonances}--\ref{fig:blowup}.

\section{Concluding remarks}
We have studied resonance and scattering from time-modulated scatterers, where the time modulation induces coupling between harmonics whose frequencies differ by multiples of the modulation frequency. The coupling terms decay exponentially, allowing small-volume asymptotic expansions and numerical calculations using  boundary integral methods. This way, we have analytically predicted, and numerically verified, a range of phenomena arising due to the time modulation: frequency conversion, (asymptotic) exceptional points, and the blowup of the scattered field at certain parameter values.  

Our methods and results open a range of future directions of study. A natural question is to study scattering from more complicated arrangements of scatterers, where the Foldy-Lax formulation can be combined with the time-modulated point-scattering approximation derived herein. For spatially periodic crystals of scatterers, a Floquet-Bloch formulation can be adopted to determine band structure and scattering from time-dependent crystals. Additionally, the boundary integral formulation can be combined with the fast multipole method \cite{greengard1987fast} to achieve fast and stable numerical methods for time-modulated problems.

\section*{Acknowledgements}

This work was partially supported by Imperial College London through a Research Impulse grant from the Cecilia Tanner Research Fund (Reference DRI251BD). B.D. was supported by the EPSRC through a fellowship with Grant No. EP/X027422/1.

\appendix
\section{One-dimensional scatterers}\label{sec:1d}
We now turn our attention to the one-dimensional case, and derive resonances of high-contrast modulated scatterers. In this case, each scatterer is an interval with time-modulated refractive index. Take a system with $N$ scatterers at centres $z_i\in \R$, separated by distances $d_{ij}$,
\begin{equation}D = \bigcup_{i=1}^N \epsilon B + z_i, \quad d_{ij} = |z_i-z_j|,\end{equation}
where $B=[-\tfrac{1}{2},\tfrac{1}{2}]$. For $i=1,...,N-1$, we let $d_i$ be the distance between two consecutive scatterers: $d_i = z_{i+1}-z_i$. We define the map $P_\epsilon: L^2(D) \to \left(L^2(B)\right)^N$
\begin{equation}P_\epsilon[\phi](x) =  \left(\begin{smallmatrix} \phi(\epsilon x+z_1)\\ \svdots \\ \phi(\epsilon x +z_N)
	\end{smallmatrix}\right).\end{equation}
In one dimension, we have the following asymptotic expansion of $\G^k$:
\begin{equation}P_\epsilon \G^k (P_\epsilon)^{-1} = \epsilon\G^{(0)} + \epsilon^2\G^{(1)} + O(\epsilon^3),\end{equation}
where $\G^{(0)}, \G^{(1)}: \left(L^2(B)\right)^N \to \left(L^2(B)\right)^N$ are block-wise defined by
\begin{equation}
	\G^{(0)}_{ij}[u_j] = \frac{e^{\iu kd_{ij}}}{2\iu k}\int_B u_j(x)\dx x,
\end{equation}
and 
\begin{equation}
	\G^{(1)}_{ij}[u_j] = \begin{cases}
		\frac{e^{\iu kd_{ij}}}{2}\int_B (x-y) u_j(x)\dx x, \quad  & i > j, \\
		-\frac{e^{\iu kd_{ij}}}{2}\int_B (x-y) u_j(x)\dx x, \quad  & i < j, \\
		\frac{1}{2}\int_B |x-y| u_j(x)\dx x, \quad  & i = j,
	\end{cases}
\end{equation}
for  $1\leq i,j \leq N.$ Following a similar analysis as in Section~\ref{sec:asymp} gives, to order $\epsilon$,
\begin{equation}\label{eq:mid}
	a_n^i + \sum_{j=1}^N \left(\frac{e^{\iu k_nd_{ij}}}{2\iu k_n } + \frac{\epsilon}{6}\delta_{ij} \right)\sum_{m=-\infty}^\infty k_m^2\epsilon \eta_{n-m}^j a_m^j = 0,
\end{equation}
for $i=1,...,N$. We consider the resonant case corresponding to the scaling $\eta = O(\epsilon^{-1})$ and define \begin{equation}
	s^i(t) = \epsilon \eta^i(t),
\end{equation}
for $s^i(t)$ independent of $\epsilon$. We define the (dense) matrix $D$ with entries $D_{ij} = e^{\iu k_nd_{ij}}$. It is straightforward to show that $D$ has a tridiagonal inverse $D^{-1} = L$ given by
\begin{equation}
	L = \left(\begin{smallmatrix} \alpha_1 & \beta_1 & &  \\
		\beta_1 & \alpha_2 & \raisebox{-4pt}{\sddots} &\\
		& \sddots & \sddots & \beta_{N-1} \\
		& & \beta_{N-1} & \alpha_N
	\end{smallmatrix}\right),
\end{equation}
where 
\begin{equation}
	\alpha_i = \begin{cases} \frac{1}{1-e^{2\iu k_n d_1}}, \quad & i = 1,\\
		\frac{1- e^{2\iu k_n(d_{i-1} + d_i)}}{(1-e^{2\iu k_n d_{i-1}})(1-e^{2\iu k_n d_i})}, & i \neq 1,N,\\
		\frac{1}{1-e^{2\iu k_n d_{N-1}}}, \quad & i = N,
	\end{cases}
\end{equation}
and
\begin{equation}
	\beta_i = \frac{e^{\iu k_n d_i}}{e^{2\iu k_n d_i} - 1}.
\end{equation}
From the Neumann series, we then have, to order $\epsilon$,
\begin{equation}
	\left(\frac{1}{\iu k_n }D + \frac{\epsilon}{3}I \right)^{-1} = \iu k_n L+ \frac{k_n^2\epsilon}{3}L^2.
\end{equation}
Multiplying \eqref{eq:mid} with this inverse, we obtain
\begin{equation}\label{eq:final1D}
	\frac{1}{2}\sum_{m=-\infty}^\infty k_m^2 S_{n-m} \mathbf{a}_m + 
	\left(\iu k_n L+ \frac{k_n^2\epsilon}{3}L^2\right)\mathbf{a}_n = 0,
\end{equation}
where $\mathbf{a}_n$ is the vector of $a_n^i$ and $S_n$ is the diagonal matrix $S_n = \mathrm{diag}\bigl(s_n^i\bigr)$.

In the case of a single scatterer (i.e. $N=1$), we  can repeat the same steps to obtain \eqref{eq:final1D} but with $L=1$.

% \begin{acknowledgments}
% B.D. acknowledges support from the Engineering and Physical Sciences Research Council (EPSRC) under grant number EP/X027422/1. The authors would like to thank Marie Touboul for insightful discussions.
% \end{acknowledgments}

\bibliography{references}

\end{document}